\documentclass[conference]{IEEEtran}
\IEEEoverridecommandlockouts
\usepackage{cite}
\usepackage{amsmath,amssymb,amsfonts}
\usepackage{algorithmic}
\usepackage{graphicx}
\usepackage{textcomp}
\usepackage{xcolor}
\usepackage{mathtools}
\usepackage{amsmath}
\usepackage{physics}
\usepackage{subcaption}
\usepackage{mathtools}
\usepackage{tablefootnote}

\usepackage{xcolor}
\usepackage{mathtools}

\usepackage{esvect}
\usepackage{gensymb}
\def\BibTeX{{\rm B\kern-.05em{\sc i\kern-.025em b}\kern-.08em
    T\kern-.1667em\lower.7ex\hbox{E}\kern-.125emX}}

\newcommand{\sigmarx}{\mbf{\Sigma}_\text{rx}}

\DeclareMathOperator{\sinc}{sinc}

\newcommand{\mbf}[1]{\mathbf{#1}}
\newcommand{\ie}{, i.e., }

\newcommand{\mym}[1]{\textbf{\textcolor{black}{#1}}}

\newenvironment{mymblock}[1]{%
    \color{black}%
}{%
    \ignorespacesafterend%
}

\begin{document}

\title{Hardware Limitations of Dynamic Metasurface Antennas in the Uplink: A Comparative Study}

\author{\IEEEauthorblockN{Maryam Rezvani\IEEEauthorrefmark{1}, Raviraj Adve\IEEEauthorrefmark{1}, Amr El-Keyi\IEEEauthorrefmark{2}, Akram bin Sediq\IEEEauthorrefmark{2}}
	\IEEEauthorblockA{\IEEEauthorrefmark{1}\textit{Dept. of Electrical and Computer Engineering},
		\textit{University of Toronto}, 
		Toronto, Canada 
	}
	\IEEEauthorblockA{\IEEEauthorrefmark{2}\textit{Ericsson Canada}, 
		Ottawa, Canada, \\
		emails: \{mrezvani, rsadve\}@ece.utoronto.ca, \{amr.el-keyi, akram.bin.sediq\}@ericsson.com}
}


\maketitle

\begin{abstract}
Dynamic Metasurface Antennas (DMAs) have emerged as promising candidates for basestation deployment in the next generation of wireless communications. While overlooking the practical and hardware limitations of DMA, previous studies have highlighted DMAs' potential to deliver high data rates while maintaining low power consumption. In this paper, we address this oversight by analyzing the impact of practical hardware limitations such as antenna efficiency, power consumed in required components, processing limitations, etc. Specifically, we investigate DMA-assisted wireless communications in the uplink and propose a model which accounts for these hardware limitations. To do so, we propose a concise model to characterize the power consumption of a DMA. For a fair assessment, we propose a wave-domain combiner, based on holography theory, to maximize the achievable sum rate of DMA-assisted antennas. We compare the achievable sum rate and energy efficiency of DMA antennas with that of a partially connected hybrid phased array. Our findings reveal the true potential of DMAs when accounting for the limitations of both designs.
\end{abstract}


\section{Introduction}
% \mym{Very short, try to squeeze it in one column}
\IEEEPARstart{H}{igh} throughput, energy efficiency, and low power consumption are key demands for the next generation of wireless communications~\cite{6Gvision}. While extra-large multiple-input multiple-output (XL-MIMO)~\cite{XLMIMO} with each antenna connected to a radio frequency (RF) chain\ie digital phased arrays (DPAs), can provide high throughput, they would result in excessive power usage at basestations (BSs). 

In contrast, hybrid MIMO antennas~\cite{HeathHYB} and dynamic metasurface antennas (DMAs)~\cite{DMA2017Analysis} connect a large number of antennas/unit-cells to fewer RF chains reducing power consumption. Specifically, while an $N$-element DPA uses $N$ RF chains, in hybrid MIMO systems\ie hybrid phased arrays (HP), using an analog combiner/precoder enables connecting $N$ antennas to $M$ RF chains, where $N \gg M$~\cite{HeathHYB}. One variant of HPs is a partially connected HP (PCHP). In PCHPs,
each antenna is connected to a phase shifter (PS), and the captured signal from $N_s$ antenna-PS pairs, is combined via a Wilkinson combiner~\cite{RiberioHYBInsertionloss} before being fed to an RF chain. %On the other hand, a fully connected HP (FCHP) employs Wilkinson power dividers and combiners along with $NM$ PSs~\cite{RiberioHYBInsertionloss}. Notably, in the context of FCHPs, a signal processing technique proposed in~\cite{SohrabiHYB} achieves the same sum rate as DPAs while adhering to an equal number of antennas constraint, provided that M in PCHP matches the number of streams served by the BS and ignoring power losses in PSs and Wilkinson power divider/combiner.

On the other hand, DMAs introduce a groundbreaking design by incorporating metasurfaces into the antenna structure, offering a wave-domain (WD) precoder/combiner. In DMAs, a large number of unit-cells are loosely coupled~\cite{DMA2017Analysis} to a few microstrip lines (or other waveguide types). Each unit-cell influences the impinging wave through a weighting factor, $w$. This factor can follow either Lorentzian phase modulation (LPM), expressed as $w=\frac{\jmath + e^{\jmath \theta}}{2}\ \theta \in [0,2\pi)$ or binary amplitude modulation (BAM), denoted as $w\in\{0,1\}$~\cite{DMAEldarUplink2019}. Notably, each unit-cell couples its signal to the neighboring microstrip line, resembling the structure of PCHPs. In DMAs, one end of each microstrip line connects to an RF chain while the other end is terminated to prevent standing waves.

Since their introduction, several aspects of DMA have been studied. For example,~\cite{DMAEldarUplink2019} studied the performance of a DMA in uplink wireless communications and showed that a LPM-constrained DMA can achieve a sum rate comparable to that of an unconstrained DMA. Furthermore, in~\cite{DMA2021Efficiency}, the authors show that a DMA offers higher energy efficiency than a DPA or fully-connected HP.

When comparing antenna types, it is crucial to consider the hardware limitations across designs. Examples of these limitations are the antenna efficiency, the power consumed by the device components, and limited beam control due to the LPM constraint. While the  DMA literature often overlooks antenna efficiency and limitations, recent work~\cite{DMA2023energy} has considered the efficiency of DMAs and the effect of sampling from the guided mode. The authors propose a tri-hybrid architecture where a DMA is used in front of a PCHP in a downlink scenario. However, the full-wave simulation used in~\cite{DMA2023energy} provides accurate modeling for the specific simulated DMA design but lacks generalizability to other DMA variants. 

Notably, the impact of DMA’s hardware limitations on uplink performance remains unexplored. In the uplink, where users contend with limited transmitted power, it is imperative to account for antenna efficiency and hardware limitations since these factors significantly impact achieving optimal performance and reliability. Equivalently, for a fair comparison, one must design effective signal processing techniques to maximize the potential performance of the chosen system.

The primary limitation of using a DMA in the uplink stems from the loose coupling between unit-cells and the underlying microstrip line~\cite{DMA2021Efficiency}. This loose coupling ensures that, in transmit mode, sampling from the guided mode and the introduced decay rate do not prevent the guided mode signal from reaching all unit-cells. The decay rate is controlled by the geometry and weighting factor of each unit-cell~\cite{DMA2021Efficiency}, respectively. However, the static control of the decay factor significantly impacts DMA performance in the uplink. It restricts the amount of power coupled to the microstrip line, subsequently affecting the signal power at the RF chain and reducing the received signal-to-noise ratio (SNR).

\textbf{Contributions}: In this paper, we aim to precisely describe the hardware limitations of DMAs. Our goal is to compare the achievable sum rate when using a DMA with when using a PCHP in uplink wireless communications. Additionally, we propose a method to account for DMA’s limitations independently of its specific design. Notably, our approach requires characterizing a single unit-cell, rather than the entire DMA structure, through full-wave simulations. We propose a model to calculate the power consumption of different designs while accounting for the power consumed by the RF chain as well as the required driving circuitry for the unit-cell's configuration. 

For the signal processing, we propose an algorithm based on holography theory to design a WD combiner which maximizes achievable sum rate in the uplink of DMA-assisted antenna systems. We compare the achievable sum rate and energy efficiency of using DMAs with using PCHP under equal aperture area and equal number of RF chains. This comparison is in single- and multi-user scenarios in rich scattering and realistic 3GPP channels while accounting for hardware limitations in both designs. Our results reveal the shortcomings of DMA-assisted antennas in comparison with using PCHP.

\textbf{Notation}: Italics denote scalars. We use lowercase and uppercase boldface letters to denote vectors and matrices, respectively. $(\cdot)^\dagger$ and $(\cdot)^H$ denote pseudo inverse and conjugate transpose, respectively. We represent the vector fields by $\vec{\mbf{A}}$. Also, $\jmath$ is defined as $\sqrt{-1}$ and $\sinc(x) = \frac{\sin(\pi x)}{\pi x}$. $\mathcal{CN}(\boldsymbol{\mu}, \mathbf{R})$ represents the complex Gaussian distribution with mean $\boldsymbol{\mu}$ and covariance matrix $\mathbf{R}$.
%%%%%%%%%%%%%%%%%%%%%%%%%%%%%%
\section{System and DMA Model}
\noindent \textbf{\textit{System Model}}: We consider a BS equipped with a DMA comprising $N_\mu$ microstrip lines with $N_e$ unit-cells on each microstrip line\ie a total number of unit-cells $N=N_\mu N_e$. The BS serves $K$ users. The number of RF chains, $M$, is equal to $N_\mu$, since each microstrip line connects to an RF chain at one end. The decoded signal at the BS is given by
\begin{equation}
	\mbf{x} = g\mbf{W}^D(\mbf{G}^\text{DMA}(\sqrt{P_T}\mbf{Hs}+\mbf{z}_\text{ant})+\mbf{z}_\text{RF})
\end{equation}
where $\mbf{x} \in \mathbb{C}^{K \times 1}$ is the received symbol vector at the BS, $P_T$ is the transmitted signal power of each user (we assume equal power transmission), $\mbf{H} \in \mathbb{C}^{N \times K}$ is channel matrix, $\mbf{z}_\text{ant} \in \mathbb{C}^{N \times 1}$ denotes the noise captured by the antenna from the environment~\cite{DMAEE}. Furthermore, $g$ and $\mbf{z}_\text{RF}\in \mathbb{C}^{M\times 1}$ are gain and noise terms introduced by the RF chain where $z_\text{RF}\sim \mathcal{CN}(0,\sigma^2_\text{RF})$. Also, via $\mbf{G}^\text{DMA} \in \mathbb{C}^{M \times N}$ and $\mbf{W}^D \in \mathbb{C}^{K \times M}$ we, respectively, model the effect of the DMA's WD and digital combiner on the received signals.

\noindent \textbf{\textit{DMA model}}: To model the DMA, we begin by pointing out its hardware limitations and the maximum weighting factor of each unit-cell.
\begin{figure}
    \centering
    \includegraphics[width = \linewidth, height=3cm]{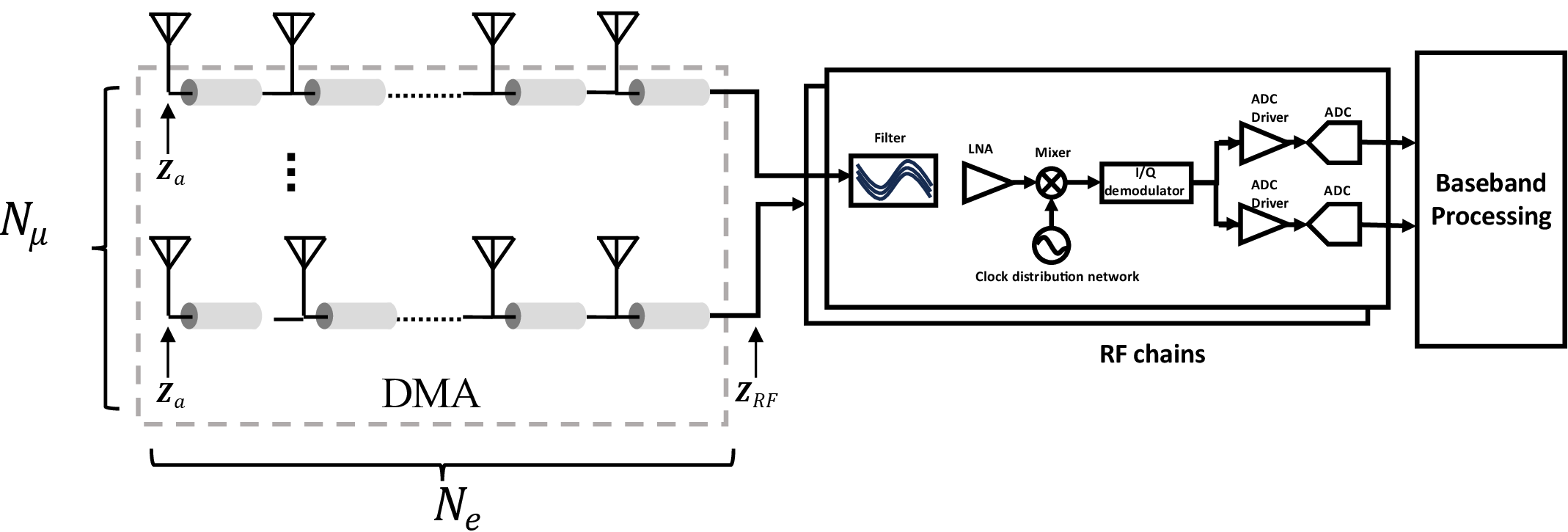}
    \caption{A schematic of a dynamic metasurface antenna}
    \label{fig:DMA}
\end{figure}
As shown in Fig.\ref{fig:DMA}, a DMA comprises $N_\mu$ microstrip lines separated by a  half-wavelength. Each line feeds the signal to (in transmit mode) or extracts the signal from (in receive mode) $N_e$ unit-cells. Each unit-cell is loosely coupled to the microstrip line to prevent extensive decay of the propagating wave in the microstrip line~\cite{2017_Smith_Diodevaractor} in transmitting mode. This requirement leads to the loss of a large portion of signal power transmitted by users in receiving mode. The coupling factor of each unit-cell is dictated by the dynamic state of the cell, average coupling factor, dielectric losses of unit-cells as well as their geometry and spacing~\cite{DMA2021Efficiency}. Therefore, while one can maximize the dynamic component of the coupling factor of each unit-cell by setting it ``on'' (in BAM mode) or by setting its weighting factor to $1\jmath$ (in LPM mode), there is a maximum coupling value dictated by the losses and geometry of the unit-cell. In what follows, we model this aspect of the coupling factor.

In a DMA, each unit-cell is modeled as a magnetic dipole with a magnetic polarizability defined by Lorentzian resonance~\cite{yoo2023uplinkDMA} described as 
\begin{equation}
    \alpha_m(\omega) = \frac{F_m\omega^2}{\omega_0^2 - \omega^2 +\jmath\Gamma_m \omega},
    \label{eq:DMALorentzianRes}
\end{equation}
where $\omega_0$ and $\omega$ are the resonance and working angular frequencies, $F_m$ is the coupling factor, and ${\Gamma}_m$ is the damping factor $=\frac{\omega_0}{2Q_m}$ where $Q_m$ is the quality factor of the unit-cell.

By ignoring the coupling between unit-cells (well-accepted as an accurate approximation~\cite{yoo2023uplinkDMA}) the magnetic dipole formed by the $i$-th unit-cell is described as
% \begin{equation}
    $\vec{\mbf{m}}_i = \alpha_{m,i}\vec{\mbf{H}}_i $
    % \label{eq:DMAm}
% \end{equation}
where $\vec{\mbf{H}}_i$ is the impinging magnetic field at the $i$-th unit-cell~\cite{DMA2017Analysis}. In the uplink, the transmitted signal by the users excites each unit-cell and hence, the $i$-th unit-cell magnetic dipole is proportional to the received  electromagnetic (EM) plane-wave described as~\cite{yoo2020DMAH}\footnote{Here we used the reciprocity theorem and described the user's transmitted EM field to be equal to the EM field radiated by the DMA as if the system is in the downlink. This ensures that the EM field coupled to the microstrip line is the dominant quasi-transverse EM (TEM) mode.}. For a single user
\begin{equation}
    \vec{\mbf{H}}_i = \sqrt{P_T}k_0^2 \eta_0 f(\vec{r})\hat{\phi}
    \label{eq:DMAH}
\end{equation}
where $P_T$ is the transmitted signal power, and $k_0$ and $\eta_0$ are free space wavenumber and impedance, respectively. In~\eqref{eq:DMAH} $f(\vec{r})$ captures the dependency of the magnetic field on the channel between the user and BS where $\vec{r}$ is their radial distance, for example in free space $f(\vec{r})= \frac{e^{-\jmath \vec{k} \cdot \vec{r}}}{4\pi \abs{\vec{r}}}$.

Combining~\eqref{eq:DMALorentzianRes}-\eqref{eq:DMAH}, the maximum magnetic dipole moment, $m^\text{DMA}_\text{max}$, occurring at the unit-cell's resonance frequency, is equal to
% \begin{equation}
    $m^\text{DMA}_\text{max} = 2 Q_m F_m k_0^2 \eta_0$.
% \end{equation}
For example, in~\cite{yoo2023uplinkDMA}, for a DMA with 30 unit-cells on each microstrip line and designed to operate at $3.5\  \text{GHz}$, $F_m = 3.0\times 10^{-9}\ [\text{m}^3]$ and $Q_m = 10.0$. In this case, the maximum amplitude of the magnetic dipole formed by each unit-cell would be $0.09$ [J/T].
which ensures the feed signal would reach all $30$ unit-cells in the underlying microstrip line. This value is consistent with the maximum value of a DMA's weighting factor in BAM mode used in~\cite{DMAEldarUplink2019}. Note that $F_m$ and $Q_m$ can be extracted from the unit-cell design using the method introduced in~\cite{UCextraction}.

Note that considering the maximum magnetic dipole moment of DMA is crucial when designing signal processing techniques or comparing it with other antenna designs. This accounts for the practicality of the proposed technique and ensures a fair comparison.

\noindent \textit{Microstrip Model}: As in~\cite{DMACE}, we capture the effect of EM field propagation in the microstrip line and weighting factor of each unit-cell in matrices $\mbf{P}^{\text{DMA}}$ and $\mbf{W}^\text{DMA}$, respectively. The DMA weighting matrix $\mbf{W}^\text{DMA}$ is given by
% \begin{equation}
    $\mbf{W}^\text{DMA} = \text{blkdiag}(\mbf{w}^1, \dots, \mbf{w}^M)$,
    % \label{eq:wDMA}
% \end{equation}
% \begin{equation}
%     \mbf{W}^\text{DMA} = 
%     \begin{bmatrix}
%         \mbf{w}^1 & \dots & 0 \\
%         \vdots & \ddots & \vdots \\
%         0 & \dots & \mbf{w}^{N_{RF}}
%     \end{bmatrix}
%     \label{eq:wDMA}
% \end{equation}
where $\mbf{W}^\text{DMA}$ is of size $N_\mu \times N$, $N=N_\mu N_e$ is the total number of unit-cells in DMA, $\mbf{w}^{i} = [w^i_1, \dots,w^i_j, \dots, w^i_{N_e}]$ is a row vector representing the weight factors of unit-cells on the $i$-th microstrip line, and $w^i_j$ is the weight factor of its $j$-th unit-cell. Each non-zero element of $\mbf{W}^\text{DMA}$ follows either a BAM or LPM constraint:
\begin{equation}
    w^{i}_{j} \in \mathcal{W}^\text{DMA} \triangleq
    \begin{cases}
        \{m^\text{DMA}_\text{max}\ \frac{\jmath+e^{\jmath\phi}}{2}, \phi \in [0,2\pi]\}& \text{LPM}\\
        \{0, m^\text{DMA}_\text{max}\} & \text{BAM.}
    \end{cases}
    \label{eq:DMAwijvalues}
\end{equation}

The propagation matrix $\mbf{P}^{\text{DMA}}$ which captures the changes in EM fields due to the propagation in the microstrip line, is described by a diagonal matrix defined as
\begin{align}
    \begin{split}
        \mbf{P}&^{\text{DMA}}_{(i-1)N_e+j} = e^{-(\alpha_w + \jmath\beta_w)d_{i,j}} ,\\
        &\forall (i,j) \in \left \{(i,j)|i \in \{1, \dots, N_\mu\}, j \in \{1, \dots, N_e\}\right \},  
    \end{split}
    \label{eq:DMAP}
\end{align}
where $d_{i,j}$ is the distance of the $j$-th element on the $i$-th microstrip line from the RF chain, $\alpha_w$ is the waveguide attenuation coefficient, and $\beta_w$ is the microstrip line wavenumber.

Combining the two matrices, the WD combiner of DMA is described as
% \begin{equation}
    $\mbf{G}^\text{DMA}= \mbf{W}^\text{DMA}\mbf{P}^\text{DMA}$.
    % \label{eq:DMAmodel}
% \end{equation}

\begin{mymblock}

\noindent \textit{Power consumption model}: The consumed power by the DMA comprises two terms related to the RF chains and the needed power to configure unit-cells. We begin with the RF chains. Accounting for the main components in RF chains\ie low noise amplifier (LNA), mixer (Mix), clock distribution network (Clk), in-phase and quadrature demodulator (IQD), analog-to-digital converter (ADC), and its driver (drv), as shown in Fig.~\ref{fig:DMA}, the power consumed, $P_{RF}$ for one RF chain is given by\footnote{A complete block diagram of RF chain can be found in~\cite{RFChain}.}
\begin{equation}
\small
    P_\text{RF} = P_\text{LNA}+P_\text{Mix}+P_\text{Clk}+P_\text{IQD}+2(P_\text{ADC}+P_\text{drv})
    \label{eq:RFchainPwr}
\end{equation}

Note that while $P_{LNA}$ is usually considered to be fixed~\cite{fixPLNA}, it contains a static term due to LNA's internal circuitry, $P_\text{sLNA}$, and a dynamic term depending on the input power, $P_\text{in}$, and power-added-efficiency, $\eta_\text{LNA}$~\cite{PAE}. Therefore, we model an LNA's power consumption as $P_\text{LNA} = \max (P_\text{sLNA}, \frac{G_{\text{LNA}}-1}{\eta_\text{LNA}}P_\text{in})$ where $G_\text{LNA}$ is the LNA gain.

To calculate the needed power to configure the DMA unit-cells, we need to carefully account for components in driving circuitry. The properties of each unit-cell are controlled via the voltage applied to the implemented varactor diode. The digital-to-analog converter (DAC) driven varactor diodes provide modulation of the unit-cell's weighting factor by translating the applied voltage to an adjustable capacitance. However, more than the power consumed by the DAC, $P_\text{DAC}$, the data line driving the DAC, $P_\text{Ctrl}$ determines the power consumed.  Finally, there is a static power consumption by the core of the DMA's field programmable gate array (FPGA), $P_\text{FPGA}$. We assume that all the processing is handled in the main FPGA/processor and the DMA's FPGA only drives the varactors (via the DAC) consuming limited power. Hence, the total power consumption of the configuration is 
\begin{equation}
    P_\text{DMA} =     N(P_\text{DAC}+P_\text{Ctrl})+P_\text{FPGA} 
    \label{eq:DMAPwr}
\end{equation}
where $N$ is the number of unit-cells. 

We note that the power consumed by the RF chains and configuring circuitry of the DMA, as defined in~\eqref{eq:RFchainPwr} and~\eqref{eq:DMAPwr} is independent of the DMA state, and only depends on the received power by the antenna via $P_\text{LNA}$.
\end{mymblock}

%%%%%%%%%%%%%%%%%%%%%%%%%%%%%%
\section{Maximizing The Achievable sum rate}

The achievable sum rate, $R$, for a DMA-assisted antenna in the uplink is defined as $R = \sum_{k=1}^K \log_2(1+\gamma_k)$ where $\gamma_k$ is the signal-to-interference-plus-noise ratio (SINR) for user $k$\ie
\begin{equation}
		\gamma_k =\frac{P_T\abs{\mbf{G}_{k,:}\mbf{h}_{:,k}}^2}{\sum_{j\neq k}^K P_T\abs{\mbf{G}_{k,:}\mbf{h}_{:,j}}^2 + z},
		\label{eq:gammaSIM}
\end{equation}
where we defined $\mbf{G} = \mbf{W}^D \mbf{G}^\text{DMA}$ and $z=\sigma_\text{ant}^2(\mbf{G}_{k,:}\mbf{\Sigma_\text{rx}}(\mbf{G}_{k,:})^H)+\sigma_\text{RF}^2 \| \mbf{W}^D_{k,:}\|^2$ represents the noise terms comprising external noise with power $\sigma_\text{ant}^2$, colored by the WD combiner, and the thermal noise added by the RF chain of power $\sigma_\text{RF}^2$~\cite{DMACE}. 

To maximize the achievable sum rate, we are interested in solving the following optimization problem:
\begin{subequations}
	\label{eq:DMAOP}
	\begin{align}
		&\max_{\mbf{W}^D, \mbf{W}^\text{DMA}} R\label{OP:obj}\\
		\text{s.t. }& \mbf{W}^\text{DMA} = \text{blkdiag}(\mbf{w}^1, \dots, \mbf{w}^M)\label{OP:cnstr1}\\
		&w^i_j \in  \mathcal{W}^\text{DMA}.\label{OP:cnstr2}
	\end{align}
\end{subequations}
where $\mathcal{W}^\text{DMA}$ is defined in~\eqref{eq:DMAwijvalues}. Solving~\eqref{eq:DMAOP} is not trivial due to the non-convexity of objective function and constraints. To obtain an effective solution, adopting holography theory~\cite{holographytheory}, we first find a desired combiner, $\mbf{G}^\text{des}$ assuming the existence of a virtual DPA at the location of the DMA. This is a trivial task using the massive MIMO literature; for example, in single-user and multi-user scenarios, a good choice for $\mbf{G}^\text{des}$ is maximum ratio combining (MRC) and zero-forcing (ZF), respectively, when noise in the system is white. Knowing $\mbf{G}^\text{des}$, we are interested in finding $\mbf{W}^\text{DMA}$ and $\mbf{W}^D$ that are solutions to the following optimization problem.
\begin{equation}
	\begin{split}
		\min_{\mbf{W}^D, \mbf{W}^\text{DMA}} &\norm{\mbf{G}^\text{des} - \mbf{W}^D \mbf{W}^\text{NW}\mbf{W}^\text{DMA}\mbf{P}^\text{DMA}}\\
		\text{s.t. } & \eqref{OP:cnstr1},\ \eqref{OP:cnstr2}
        \label{eq:DMAOP2}
	\end{split}   
\end{equation}
where, since $\mbf{G}^\text{des}$ is a good choice for systems with added white noise and in DMA we are dealing with colored noise, $\mbf{W}^\text{NW} = \mbf{\Sigma}_z^{-\frac{1}{2}}$ is introduced as a noise whitening filter where $\mbf{\Sigma}_z$ is the noise correlation matrix.
While~\eqref{eq:DMAOP2} is still non-convex jointly in $\mbf{W}^\text{D}$ and $\mbf{W}^\text{DMA}$, adopting an alternative optimization approach, similar to~\cite{DMACE}, we find an optimum answer for $\mbf{W}^D$ and $\mbf{W}^\text{DMA}$. 

\noindent \textit{WD combiner}: Given $\mbf{W}^D$, various solutions have been proposed to obtain $\mbf{W}^\text{DMA}$. Here, we use the non-zero mapper (NZM) and the closed-form mapper (CFM) as introduced in~\cite{DMAEldarUplink2019} and~\cite{DMACE}. Specifically, defining $\mbf{G}^D = \mbf{W}^D \mbf{W}^\text{NW}$, NZM finds the closest WD combiner to $\mbf{G}^\text{des}$ by normalizing the Frobenious norm between $(\mbf{G}^D)^\dagger\mbf{G}^\text{des}(\mbf{P}^\text{DMA})^{-1}$ and $\mbf{W}^\text{DMA}$ entry-wise. On the other hand, in CFM, knowing $\mbf{G}^D$ a closed-form solution for elements of $\mbf{W}^\text{DMA}$ that minimizes Frobenious distance between $\mbf{G} = \mbf{G}^\text{des}(\mbf{P}^\text{DMA})^{-1}$ and $\mbf{G}^D \mbf{W}^\text{DMA}$ is calculated as $w^i_j = \frac{(\mbf{G}^D)_i^H (\mbf{G})_{i,j}}{\|(\mbf{G}^D)_i\|^2}$ where $(\mbf{G})_{i,j}$ and $\mbf{G}^D_i$ denote the $j$-th column of the $i$-th $M\times N_e$ block of $\mbf{G}$ and $i$-th column of $\mbf{G}^D$. Then $w^i_j$ is first mapped to the unit-circle followed by projection to the Lorentzian circle to obtain its final value.

Conversely, given $\mbf{W}^\text{DMA}$ to calculate the digital combiner, we adopt the following approaches:
\begin{itemize}
    \item \textit{Single-user case}: In this case, with $\mbf{W}^\text{DMA}$ known, the best solution is applying a noise whitening filter $\mbf{W}^\text{NW}$ followed by match filtering defined as $\mbf{W}^D = (\mbf{W}^\text{NW} \mbf{W}^\text{DMA} \mbf{P}^\text{DMA} \mbf{H})^H$
    \item \textit{Multi-user case}: Knowing $\mbf{W}^\text{DMA}$, an optimal choice for $\mbf{W}^D$ is the least-squares solution\ie $\mbf{W}^D = \mbf{G}^\text{des}(\mbf{W}^\text{NW}\mbf{W}^\text{DMA}\mbf{P}^\text{DMA})^\dagger$.
\end{itemize}

The process of alternatively optimizing between $\mbf{W}^\text{DMA}$ and $\mbf{W}^D$ continues until convergence.

\section{Numerical Study}

\begin{figure*}[t]
	\begin{subfigure}{0.25\textwidth}
		%		 \begin{minipage\right }[b]{0.49\linewidth}
			\centering
			\includegraphics[width=0.9\linewidth]{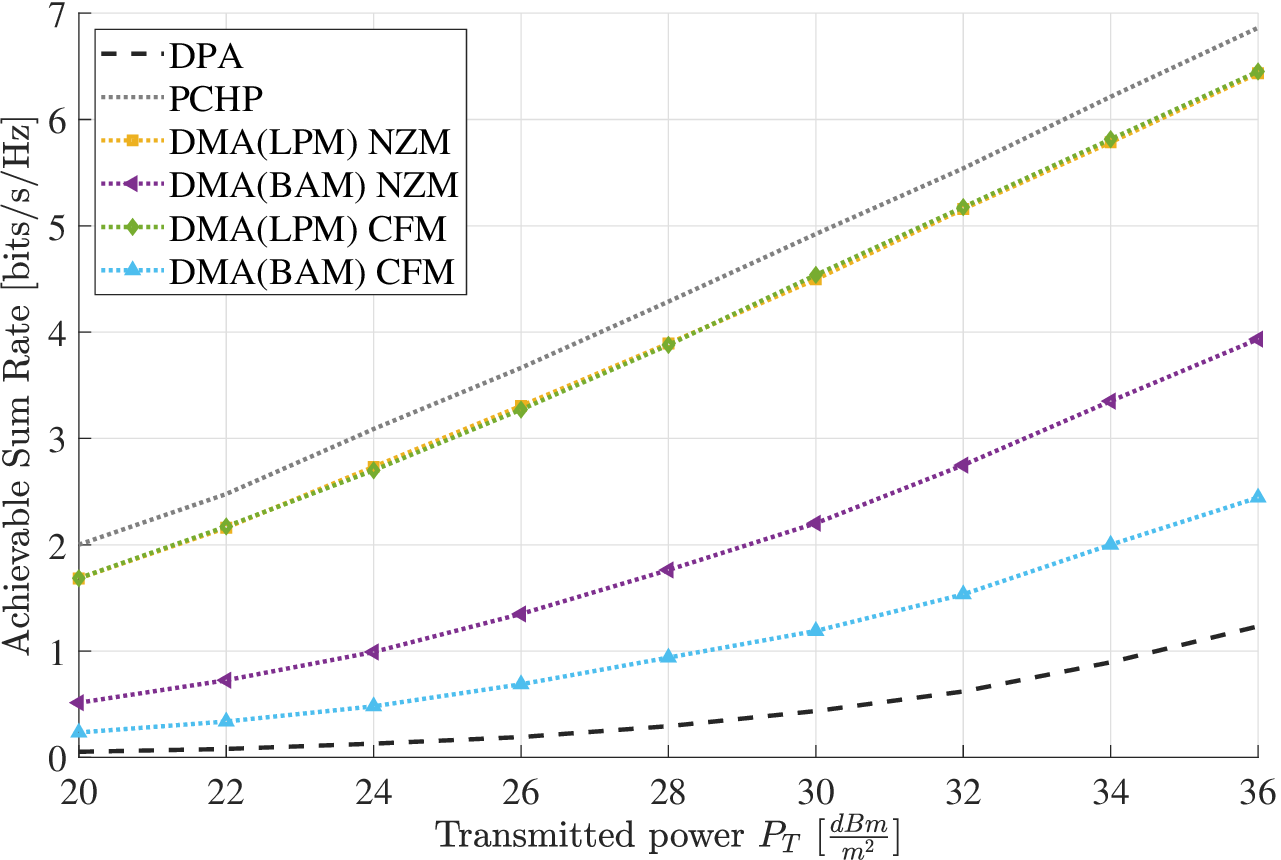}
			% \subcaption{(b)}
			\caption{}
			\label{fig:SURNL}
			% \end{minipage}
	\end{subfigure}
	\hfill
	\begin{subfigure}{0.25\textwidth}
		% \begin{minipage}[b]{0.49\linewidth}
			\centering
			\includegraphics[width=0.9\linewidth]{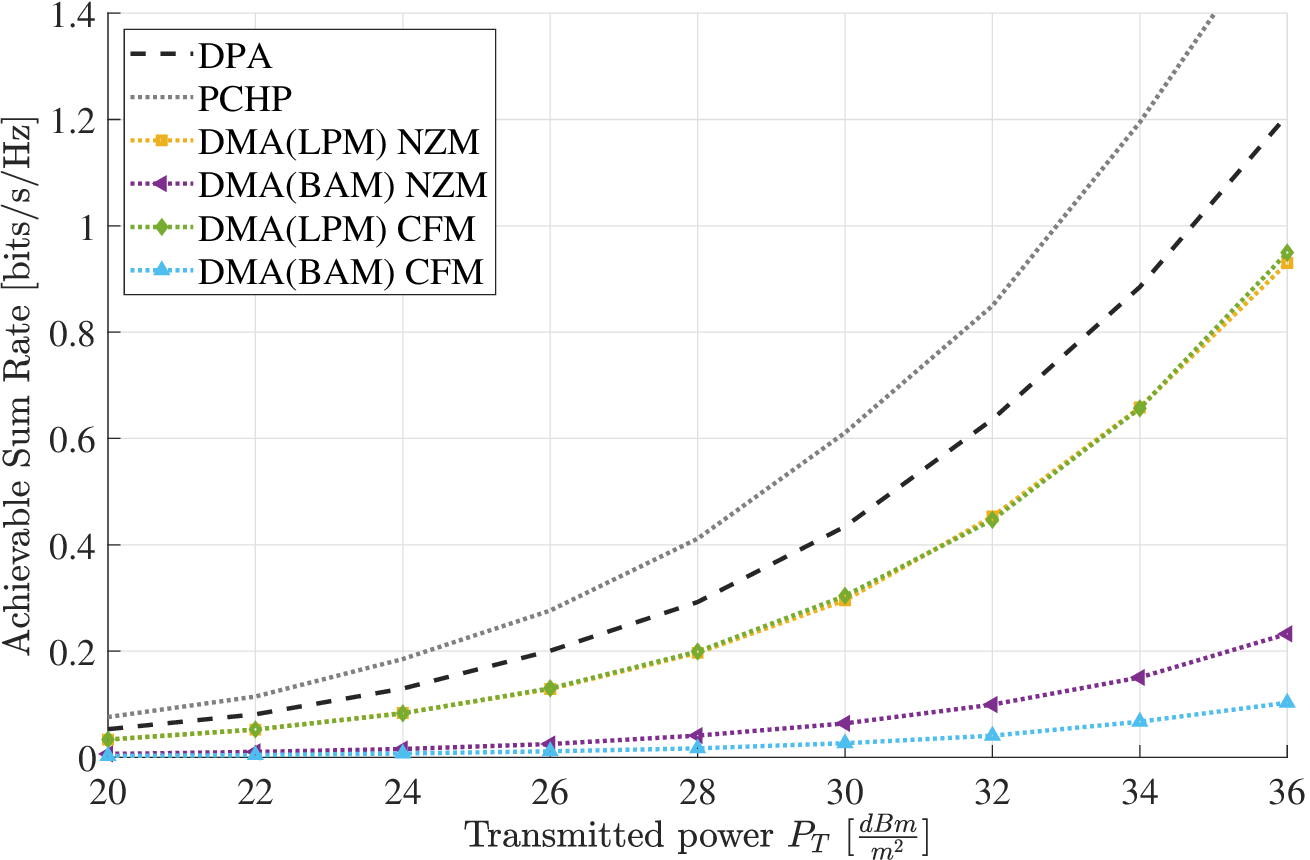} 
			\caption{}
			\label{fig:SURL}
			% \end{minipage}
	\end{subfigure}
 \hfill
	\begin{subfigure}{0.25\textwidth}
		% \begin{minipage}[b]{0.49\linewidth}
			\centering
			\includegraphics[width=0.9\linewidth]{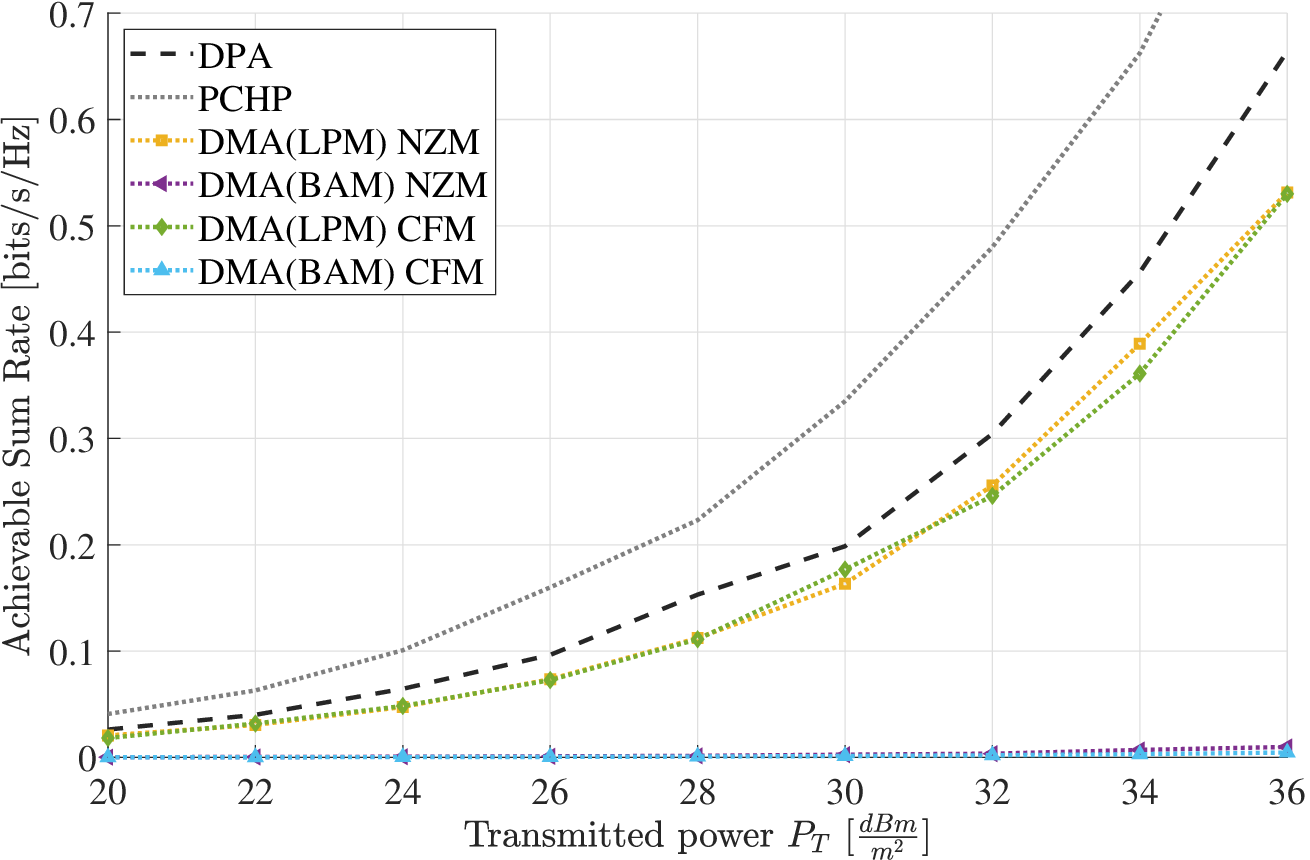} 
			\caption{}
			\label{fig:SUQ}
			% \end{minipage}
	\end{subfigure}
	\caption{$K=1,M=4$. Achievable sum rate (a) ignoring, (b) factoring hardware limitations in rich scattering environment. (c) Achievable sum rate in realistic 3GPP channels factoring hardware limitations.  }
	\label{fig:SESU}
\end{figure*}
\begin{figure*}[!t]
	\begin{subfigure}{0.25\textwidth}
		%		 \begin{minipage\right }[b]{0.49\linewidth}
			\centering
			\includegraphics[width=0.9\linewidth]{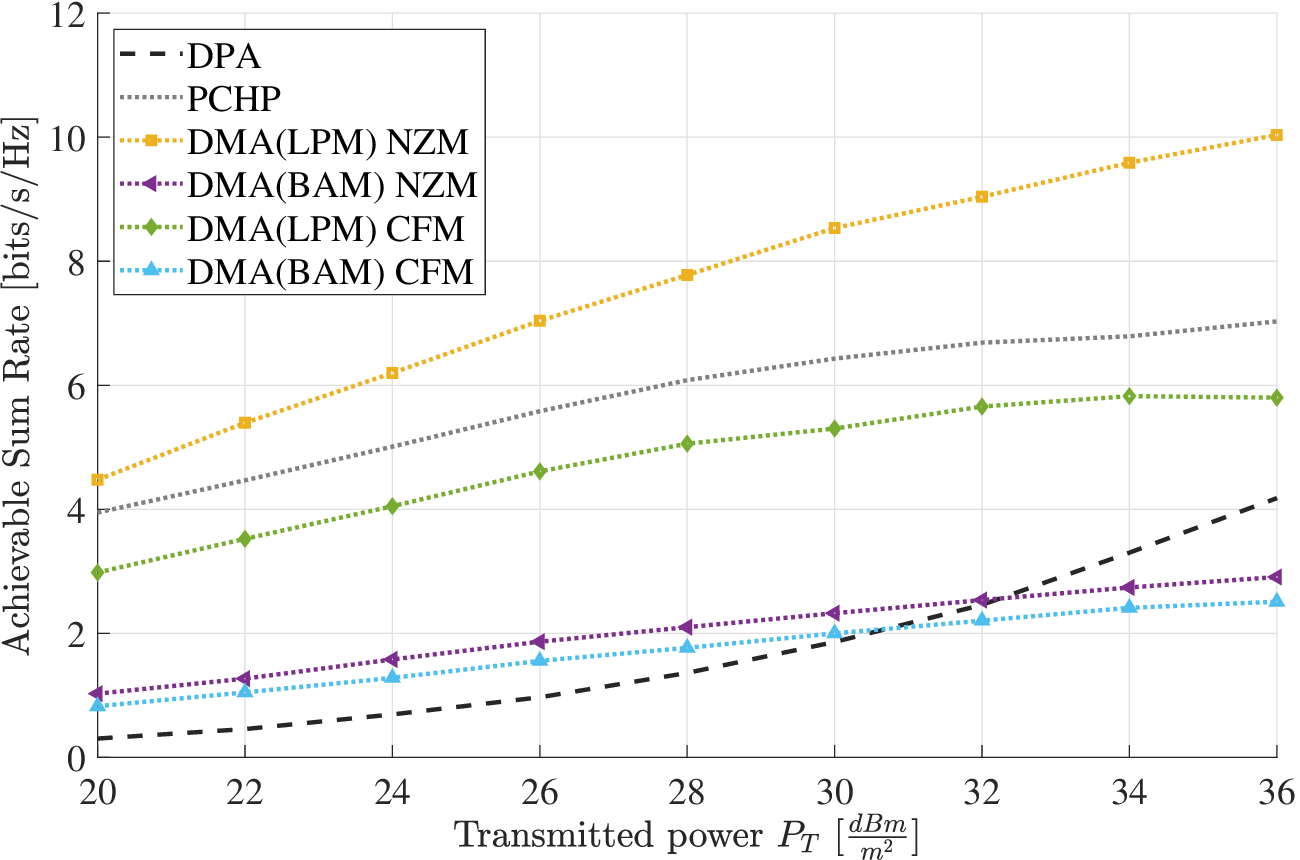}
			% \subcaption{(b)}
			\caption{}
			\label{fig:MURNL}
			% \end{minipage}
	\end{subfigure}
	\hfill
	\begin{subfigure}{0.25\textwidth}
		% \begin{minipage}[b]{0.49\linewidth}
			\centering
			\includegraphics[width=0.9\linewidth]{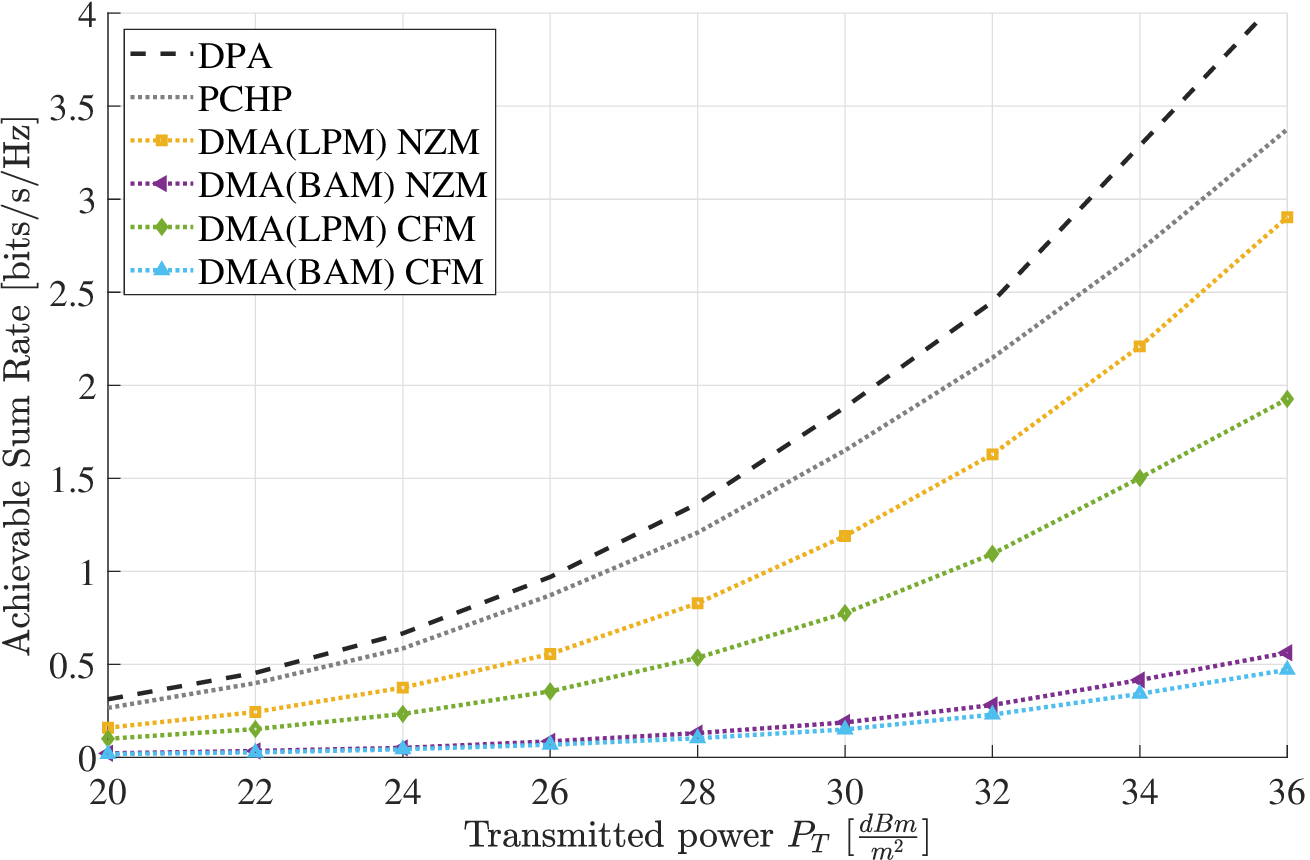} 
			\caption{}
			\label{fig:MURL}
			% \end{minipage}
	\end{subfigure}
        \hfill
	\begin{subfigure}{0.25\textwidth}
		% \begin{minipage}[b]{0.49\linewidth}
			\centering
			\includegraphics[width=0.9\linewidth]{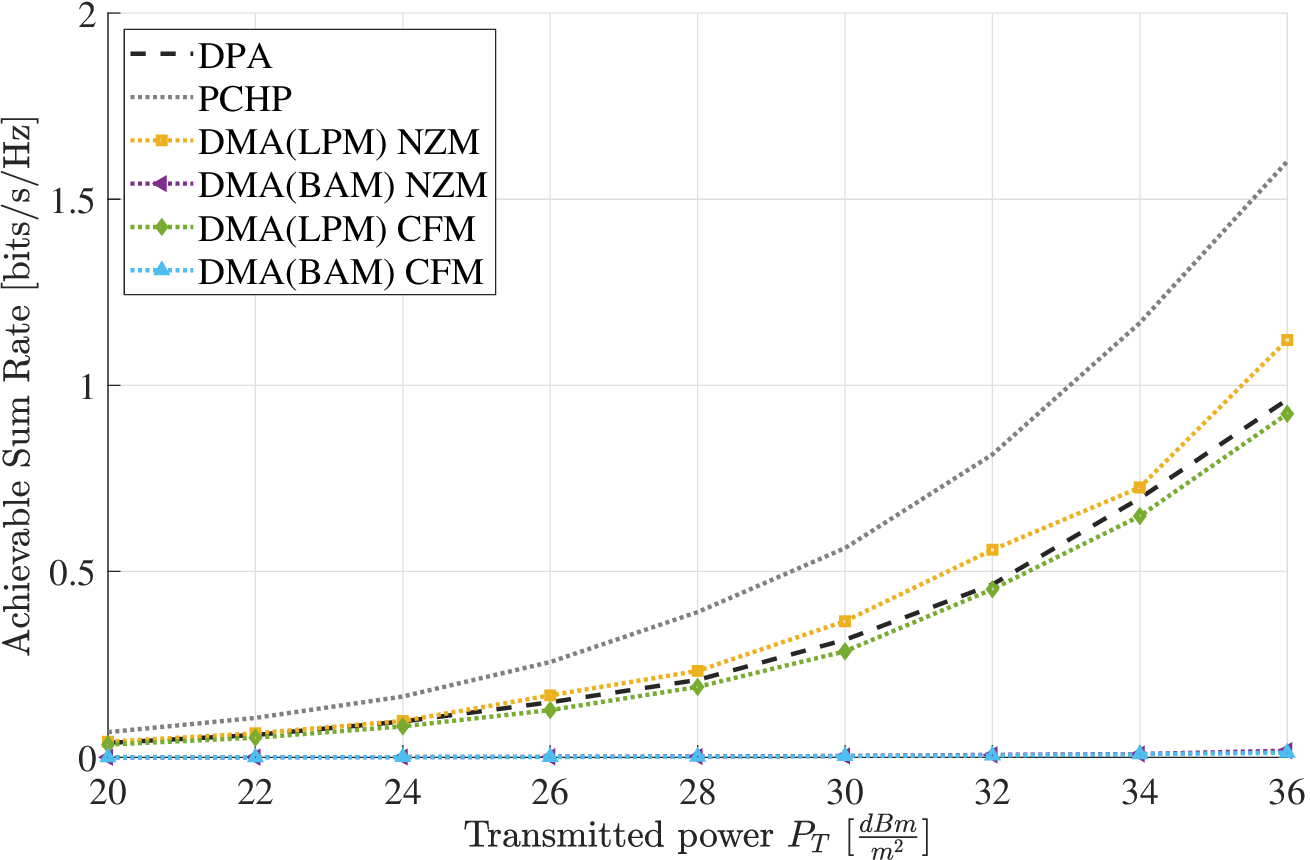} 
			\caption{}
			\label{fig:MUQ}
			% \end{minipage}
	\end{subfigure}
	\caption{$K=2,M=4$. Achievable sum rate (a) ignoring (b) factoring hardware limitations in rich scattering environment. (c) Achievable sum rate in realistic 3GPP channels factoring hardware limitations. }
	\label{fig:SEMU}
\end{figure*}
In this section, we compare the sum rate when using a DMA with using a PCHP, considering the hardware issues described above. We examine equal aperture areas, RF chains $M$, single-user and multi-user scenarios, in an outdoor wireless communication system where $K$ users communicate with a base station at $3$ GHz carrier frequency with $20$ MHz bandwidth. As the BS is in the far-field of users, it receives a plane wave with Poynting vector magnitude equal to $P_T  \frac{\abs{\vec{\mbf{E}}_\text{user}}^2}{2\eta_0}$ $[\frac{W}{m^2}]$ where $P_T$ is the power transmitted by users. 

We assume both PCHP and DMA have physical apertures equal to $2\lambda_0 \times 8\lambda_0$ and $M=4$ RF chains. The PCHP is an array of patch antennas separated by $\lambda_0/2$ with $90\%$ efficiency~\cite{Patchantenna}. Also, the unit-cells of the DMA are of size $\lambda_0/2 \times \lambda_0/6$~\cite{DMA2018SmithPINpowerConsumptionCombiner}. This configuration translates to $N_s$ and $N_e$ equal to $16$ and $48$, respectively. Furthermore, the RF chain is at room temperature\ie $T_\text{BS} = 290 \degree\ [K]$. For a microstrip line as described in~\cite{yoo2023uplinkDMA} $\alpha_w$ and $\beta_w$ are set $0.13$ and $113.8\ [\text{m}^{-1}]$, respectively. Finally, $g$ and the noise figure of the RF chain are set to be $12.5$ [dB] and $18.8$ [dB], respectively. Also, to form a clear picture of the effect of hardware limitations, we report the performance of DPA with $N = M=4$. We repeat the simulation for 10 instances of user placements and 100 channel realizations per instance.

The power captured by each antenna is equal to $P_T A_\text{eff}$ where $A_\text{eff}$ is the antenna effective area~\cite[Ch.~2]{balanis2016antenna}. Each unit-cell is approximated by a continuous current sheet, hence, $A_\text{eff,UC}$ is equal to its physical area~\cite[Ch.~2]{balanis2016antenna}. In the case of PA, using Matlab's Antenna Designer app~\cite{MATLAB}, a typical patch antenna is designed and its effective area is calculated.

The noise power captured by the antenna is given by 
%\begin{equation}
$\sigma_\text{ant}^2 = KT_EB\frac{A_\text{eff}}{\lambda_0^2}\Delta \Omega$~\cite[Ch.~13]{bhattacharyya2006phased} 
%\end{equation}
where $K$, $T_E$, and $\Delta\Omega$ are Boltzmann's constant, environment temperature ($290 \degree \ K$)~\cite{Stutzman}, and the solid angle of noise sources equal to $2 \pi$.

For rich scattering environments, we use the Kronecker Rayleigh fading model where $\mbf{H} = \mbf{\Sigma}_{rx}\hat{\mbf{H}}\mbf{\Sigma}_{tx}$, $\hat{\mbf{H}} \sim \mathcal{CN}(\mbf{0},\mbf{I}_{NK})$, and $\mbf{\Sigma}_{tx}$ and $\mbf{\Sigma}_{rx}$ are transmitters'\ie users', and receivers'\ie antennas', correlation matrices, respectively. We assume independent channel between users\ie $\mbf{\Sigma}_{tx} = \text{diag}(\beta_1, \dots, \beta_K)$ where $\beta_k$ is channel path loss of $k$-th user computed using the Winner II urban microcell (UMi) non-line of sight (NLoS) scenario~\cite{Winner}. To calculate $\mbf{\Sigma}_{rx}$, we assume scatterers are distributed in 3D space and hence, $(\sigmarx)_{(i,j)} = \sinc(\frac{2d_{i,j}}{\lambda_0})$~\cite{paulraj2003introduction} where $d_{i,j}$ is the distance between $i$-th and $j$-th unit-cells. Also, to evaluate the achievable sum rate of different antenna designs in realistic 3GPP channels, we use channels generated by QuadRiGa~\cite{quadriga}. In realistic channels the external noise correlation matrix is equal to that in a rich scattering environment\ie $\mbf{z}_\text{ant}\sim \mathcal(0,\sigma_\text{ant}^2 \sigmarx )$.

In single-user and multi-user scenarios, MRC and ZF combiners are used for the DPA. The PCHP's combiner is obtained using the algorithm in~\cite{MOHYB}. Finally, to account for hardware limitations in PCHP, we assume a $k$-way Wilkinson combiner is formed by cascading $\lceil \frac{N}{M} \rceil$ $2$-way Wilkinson combiners~\cite{RiberioHYBInsertionloss}. The insertion loss of the Wilkinson power combiner and phase shifter (PS) is $3.9\ [\text{dB}]$~\cite{wilkinson} and $5\ [\text{dB}]$~\cite{PSpassive}, respectively.

\noindent \textit{Single-user scenario}: Depicted in Fig.~\ref{fig:SESU} is the achievable sum rate of different designs in the single-user case. As shown in Fig.~\ref{fig:SURNL}, when hardware limitations are not factored in, the PCHP and DMA achievable sum rate surpasses that of DPA with equal $M$. In PCHP, PSs can fully compensate for the phase differences in signal captured by different antennas, while in DMA, due to LPM (or BAM) constraints, signals captured by different unit-cells are not completely in phase. This explains the performance gap between DMA and PCHP. In contrast, when hardware limitations are accounted for in rich scattering channels, as illustrated in Fig.~\ref{fig:SURL}, the achievable sum rate of the PCHP is close to that of the DPA and the DMA's performance drops dramatically. The superior performance of PCHP with respect to DPA is mainly due to its larger physical aperture. which leads to higher signal power captured from the environment. 

Due to the small coupling factor between unit-cells and microstrip lines in the DMA structure, most of the signal power illuminating the antenna's aperture is not coupled to microstrip lines leading to poor performance of the DMA in comparison to DPA and PCHP.

Finally, the achievable sum rate of different designs in realistic 3GPP channels is demonstrated in Fig.~\ref{fig:SUQ}. In these selective channels, while the performance of all designs drops, DPA experiences the most degradation in its achievable sum rate due to its smaller aperture area. Furthermore, the achievable sum rate of the BAM-constrained DMA case is extremely low which makes this design inappropriate.
\begin{figure}[t]
	\begin{subfigure}{0.20\textwidth}
		%		 \begin{minipage\right }[b]{0.49\linewidth}
			\centering
			\includegraphics[width=0.9\linewidth]{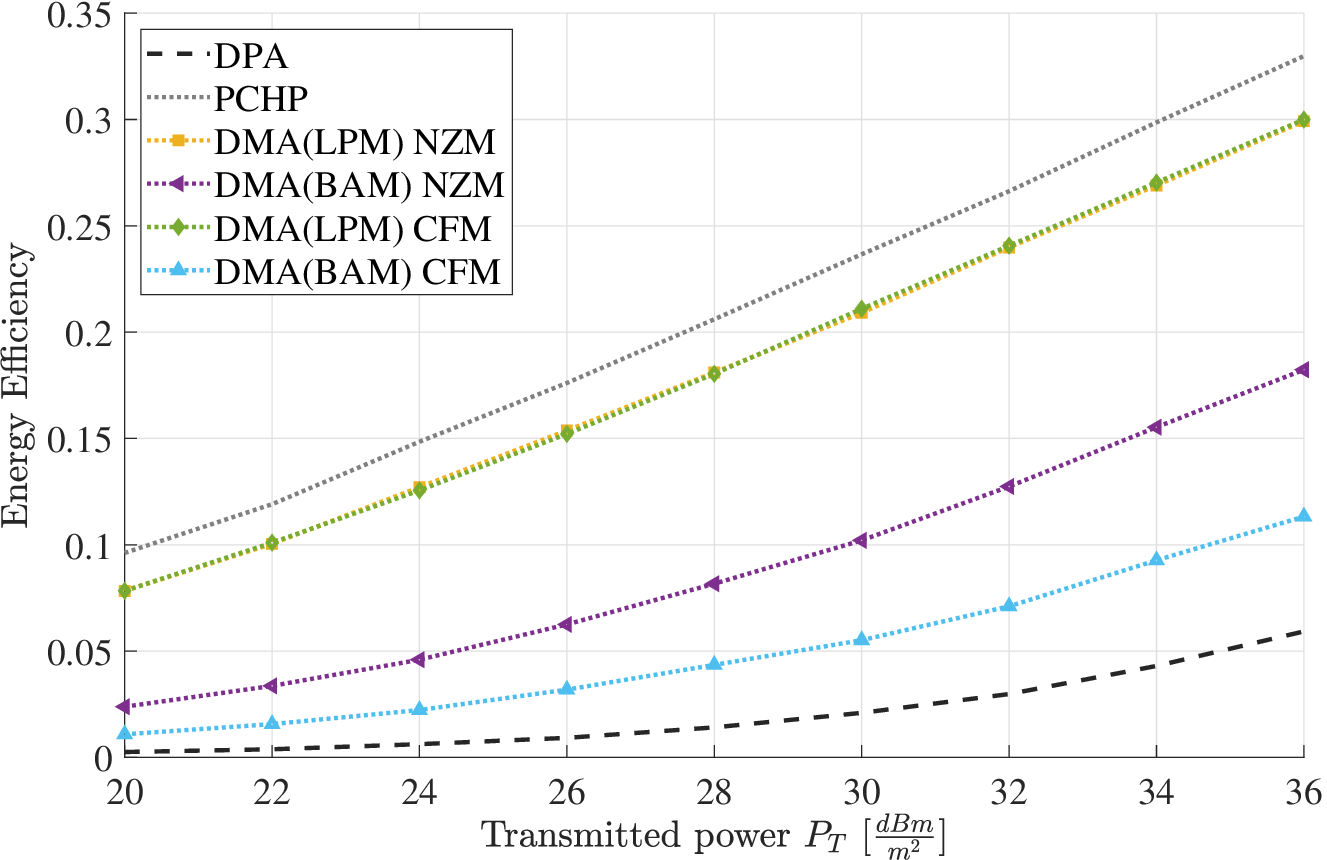}
			% \subcaption{(b)}
			\caption{}
			\label{fig:ESURNL}
			% \end{minipage}
	\end{subfigure}
	\hfill
	\begin{subfigure}{0.20\textwidth}
		% \begin{minipage}[b]{0.49\linewidth}
			\centering
			\includegraphics[width=0.9\linewidth]{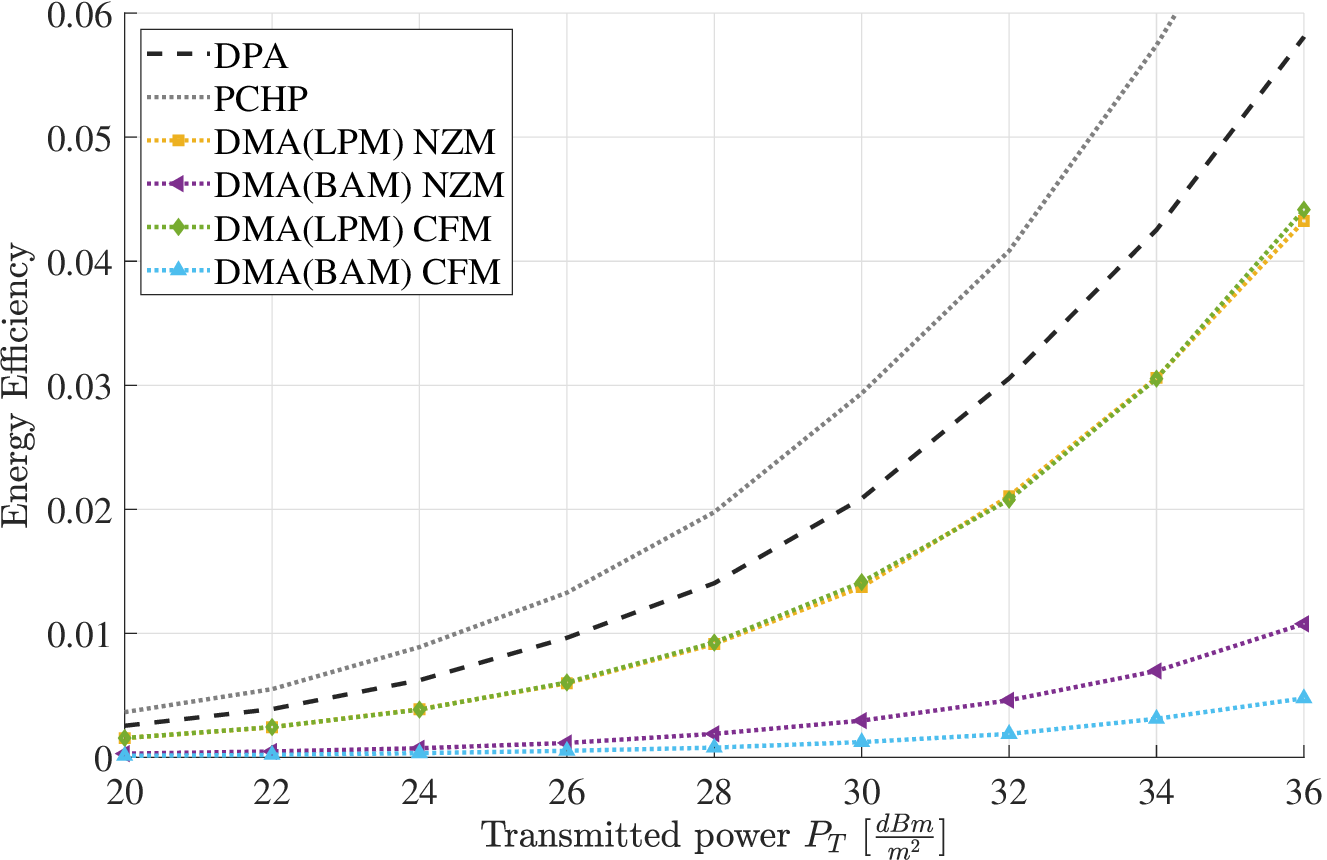} 
			\caption{}
			\label{fig:ESURL}
			% \end{minipage}
	\end{subfigure}
 \hfill
        \begin{subfigure}{0.20\textwidth}
		%		 \begin{minipage\right }[b]{0.49\linewidth}
			\centering
			\includegraphics[width=0.9\linewidth]{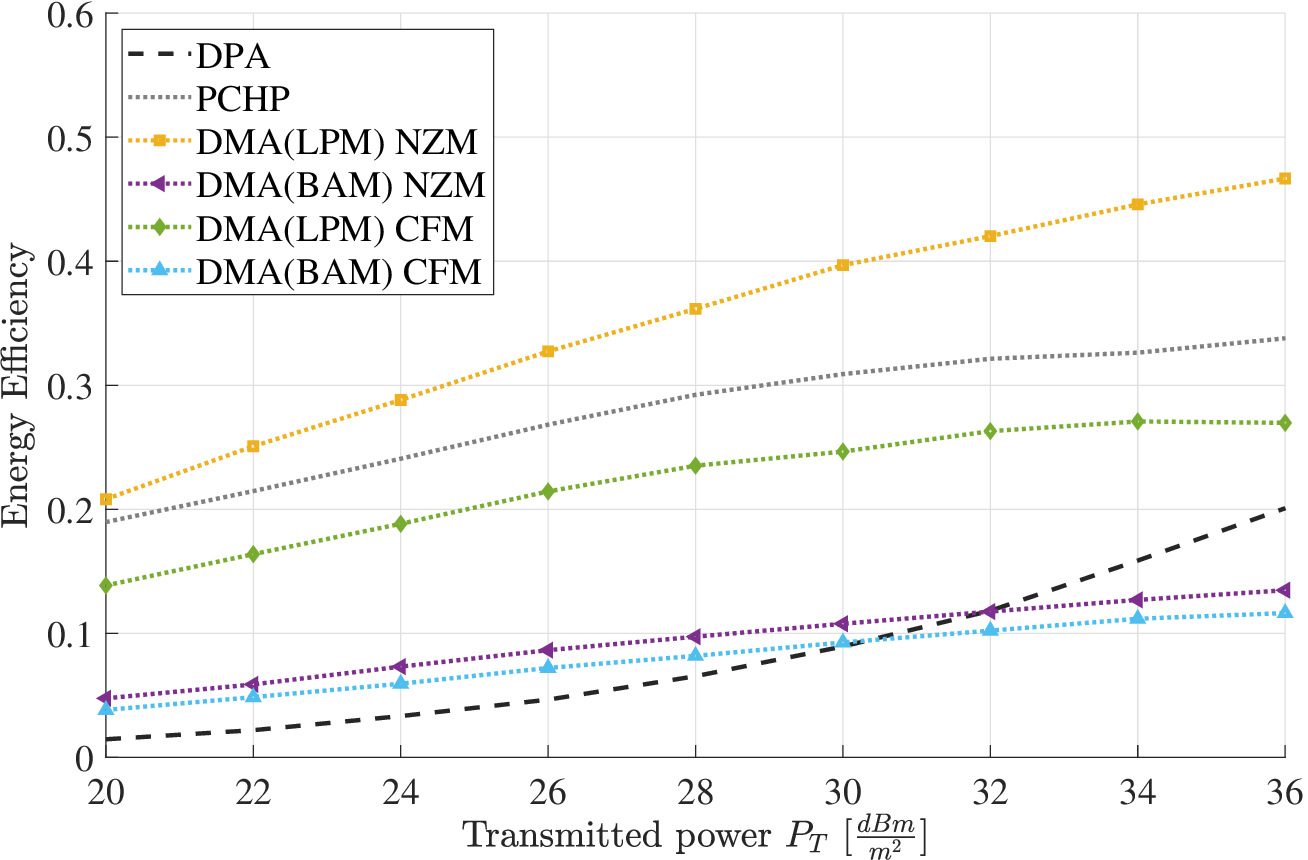}
			% \subcaption{(b)}
			\caption{}
			\label{fig:EEMURNL}
			% \end{minipage}
	\end{subfigure}
	\hfill
	\begin{subfigure}{0.20\textwidth}
		% \begin{minipage}[b]{0.49\linewidth}
			\centering
			\includegraphics[width=0.9\linewidth]{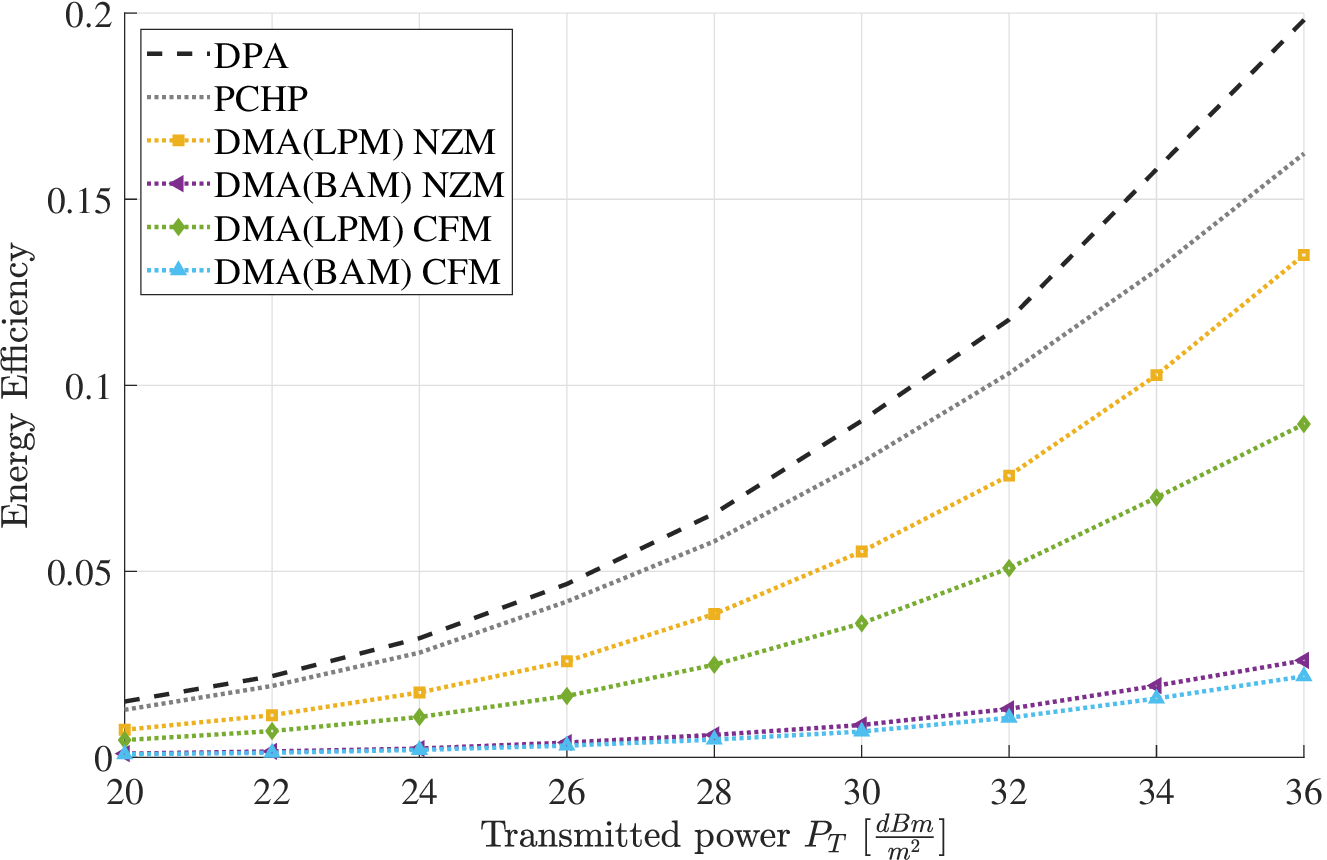} 
			\caption{}
			\label{fig:EEMURL}
			% \end{minipage}
	\end{subfigure}
	\caption{Energy efficiency  in rich scattering environments: single-user scenario (a) ignoring, (b) factoring hardware limitations, multi-user scenario (b) ignoring, (c) factoring hardware limitations.  }
	\label{fig:EE}
\end{figure}

\noindent \textit{Multi-user scenario}: The performance of different antenna designs in a multi-user scenario where BS serves $2$ users is depicted in Fig.~\ref{fig:SEMU}. Similar to the single-user case, when hardware limitations are not accounted for, as shown in Fig.~\ref{fig:MURNL}, both PCHP and the DMA achieve higher sum rates than the DPA due to their larger number of antennas and available flexibility to form more complex patterns. Also, the DMA's performance in LPM mode using NZM surpasses that of PCHP mainly due to dense unit-cells in the DMA structure and the flexibility it provides to nullify inter-user interference. Moreover, the achievable sum rate of the DMA in BAM mode has the same trajectory as that of PCHP. In contrast, when accounting for hardware limitations, in rich scattering channels, the DPA's achievable sum rate surpasses other designs due to its lower signal losses and, hence, higher SINR. On the other hand, in realistic 3GPP channels, PCHP's performance is better than the DPA due to its larger effective aperture and increased signal power. Here, the DMA's sum rate is close to that of the DPA, due to its large aperture area but lower than PCHP due to higher losses in the system. 

\noindent \textit{Energy Efficiency}: Energy efficiency (EE) is defined as $\text{EE} = \frac{R}{P}$. The power consumption of the DPA and PCHP is due to $M$ RF chains calculated using~\eqref{eq:RFchainPwr}. In contrast, DMA power consumption is due to RF chains and unit-cell's configuring circuitry as described in~\eqref{eq:DMAPwr}. To calculate consumed power in each case, $P_\text{sLNA}$~\cite{LNA}, $P_\text{Mix}$~\cite{mixer}, $P_\text{IQD}$~\cite{IQdem}, $P_\text{drv}$~\cite{ADCdrv}, $P_\text{ADC}$~\cite{ADC}, $P_\text{Clk}$~\cite{CLK}, $P_\text{DAC}$~\cite{DAC}, $P_\text{Ctrl}$~\cite{AMDxilinx} and $P_\text{FPGA}$~\cite{AMDxilinx} are set as $0.75$ [W], $0.4$ [W], $2.2$ [W], $0.15$ [W], $0.725$ [W], $0.1$ [W], $0.002$ [W], $0.0006$ [W], and $0.1$ [W], respectively. Finally, $\eta_\text{LNA}$ is 0.12~\cite{LNA}. Note that in calculating $P_\text{LNA}$, we assume the power received by each antenna is equal to $\beta_A P_T A_{\text{eff}} $ where $\beta_A$ is the average path-loss of the users. Also, in a DMA/PCHP, the total power captured by $N_s$/$N_e$ antennas enters each LNA while in DPA, the power captured by one antenna forms $P_{in}$ of the LNA.
%     \begin{table}[h]
%     \centering
%     \begin{tabular}{|c|c|}
%         \hline
%         Device & Power [W]\\
%         \hline
%         $P_{\textbf{LNA}}$~\cite{LNA} & 0.75\\
%         \hline
%         $P_{Mix}$~\cite{mixer} & 0.4\\
%         \hline
%         $P_{IQD}$~\cite{IQdem}& 2.2\\
%         \hline
%         $P_{drvADC}$~\cite{ADCdrv}& 0.15\\
%         \hline
%         $P_{ADC}$~\cite{ADC}& 0.725\tablefootnote{Calculated using information provided in~\cite{ADCNF} and signal-to-noise-and-distortion ratio of ADC}\\
%         \hline
%         $P_{Clk}$~\cite{CLK}& 0.1\\
%         \hline
%         $P_{DAC}$~\cite{DAC}& 0.002\\
%         \hline
%         $P_{Ctrl}$~\cite{AMDxilinx}& 0.0006\\
%         \hline
%         $P_{FPGA}$~\cite{AMDxilinx}& 0.1\\
%         \hline
%     \end{tabular}
%     \caption{Power consumption of different devices}
%     \label{tab:Pwr}
% \end{table}

Depicted in Fig.~\ref{fig:EE}, is the EE in rich scattering channels. As seen, the EE shows the same trend as the achievable sum rate. This is because the major contributor to the power consumption of the antenna is the number of RF chains, $M$, which is equal across all compared designs. Also, the DMA's power consumption is slightly larger than the other two antenna designs due to the required circuitry to configure the state of unit-cells. Note that combining a DMA with HPs, as in~\cite{DMA2023energy}, would result in a more energy-efficient design. 
% \begin{figure*}[!t]
% 	\begin{subfigure}{0.32\textwidth}
% 		%		 \begin{minipage\right }[b]{0.49\linewidth}
% 			\centering
% 			\includegraphics[width=0.9\linewidth]{Figures/AggregateEE_DMA_Noloss_Final_Rayleigh_B1_PL_sizex8_sizey2_M4_Mx1_K2_GLNA15_L1_HMSfeed1.eps}
% 			% \subcaption{(b)}
% 			\caption{}
% 			\label{fig:EEMURNL}
% 			% \end{minipage}
% 	\end{subfigure}
% 	\hfill
% 	\begin{subfigure}{0.32\textwidth}
% 		% \begin{minipage}[b]{0.49\linewidth}
% 			\centering
% 			\includegraphics[width=0.9\linewidth]{Figures/AggregateEE_DMA_loss_Final_Rayleigh_B1_PL_sizex8_sizey2_M4_Mx1_K2_GLNA15_L1_HMSfeed1.eps} 
% 			\caption{}
% 			\label{fig:EEMURL}
% 			% \end{minipage}
% 	\end{subfigure}
%         \hfill
% 	\begin{subfigure}{0.32\textwidth}
% 		% \begin{minipage}[b]{0.49\linewidth}
% 			\centering
% 			\includegraphics[width=0.9\linewidth]{Figures/AggregateEE_DMA_loss_Final_QuadRiGA_B1_PL_sizex8_sizey2_M4_Mx1_K2_GLNA15_L1_HMSfeed1.eps} 
% 			\caption{}
% 			\label{fig:EEMUQ}
% 			% \end{minipage}
% 	\end{subfigure}
% 	\caption{$K=2,M=4$. Energy efficiency (a) ignoring (b) accounting for hardware limitations in rich scattering environment. (c) Energy efficiency in realistic 3GPP channels accounting for hardware limitations. }
% 	\label{fig:EEMU}
% \end{figure*}

\section{Conclusion}
In this paper, we studied the impact of the hardware limitations of a DMA in uplink wireless communications. Our results show that the hardware limitations of DMA drastically affect its achievable sum rate in comparison with PCHP and DPA with an equal number of RF chains. This underlines the importance of accounting for the practical limitations of DMA, indeed any other antenna design, when used in a wireless communication scenario.

\section*{Acknowledgments}
%%%%%%%%%%%%%%%%%%%%%%%%%%%%%%%%%%%%%%%%%%%%%%%%%%%%%%%%%%%%%%%%%%%%%%%%%%%%%%%%%%%%%
This work was supported by the Ericsson Canada. 
% Maryam Rezvani gratefully acknowledges insightful discussions with Parham Abbasloo.
%%%%%%%%%%%%%%%%%%%%%%%%%%%%%%%%%%
\bibliographystyle{IEEEtran}
\bibliography{references}

\end{document}